\documentclass{article}

\PassOptionsToPackage{numbers, compress}{natbib}
\usepackage[final]{tackling_climate_workshop_style}
\newcommand{\AI}{AI2 }  

\usepackage[utf8]{inputenc} 
\usepackage[T1]{fontenc}    
\usepackage{hyperref}       
\usepackage{url}            
\usepackage{booktabs}       
\usepackage{amsfonts}       
\usepackage{nicefrac}       
\usepackage{microtype}      
\usepackage{xcolor}
\usepackage{graphicx}
\usepackage{xfrac}

\RequirePackage{doi}

\usepackage{color, colortbl}
\definecolor{darkblue}{rgb}{0.0,0.0,0.65}
\definecolor{darkred}{rgb}{0.68,0.05,0.0}
\definecolor{darkgreen}{rgb}{0.0,0.29,0.29}
\definecolor{darkpurple}{rgb}{0.47,0.09,0.29}
\definecolor{myblue}{rgb}{0.1380392156862745, 0.3027450980392157, 0.6274509803921569}
\definecolor{myred}{rgb}{0.6235294117647059, 0.13725490196078433, 0.0196078431372549} 
\hypersetup{
  colorlinks = true,
  citecolor  = myblue,
  linkcolor  = myred,
  filecolor  = darkblue,
  urlcolor   = darkblue,
 }

\usepackage{tikz, }
\usetikzlibrary{shapes,arrows,fit}

\title{ACE: A fast, skillful learned global atmospheric model for climate prediction}

\author{%
  Oliver Watt-Meyer\thanks{\texttt{oliverwm@allenai.org}}, Gideon Dresdner, Jeremy McGibbon, Spencer K. Clark, Brian Henn, \\
  {\bf Matthew E. Peters, and Christopher S. Bretherton}\\
  Allen Institute for Artificial Intelligence (AI2), Seattle, USA
  \And
  James Duncan\thanks{2023 summer intern at AI2} \\
  Graduate Group in Biostatistics, University of California, Berkeley USA \\
  \And
  Noah D. Brenowitz, Karthik Kashinath, Michael S. Pritchard, and Boris Bonev \\
  NVIDIA, Santa Clara, USA
}

\begin{document}

\maketitle

\begin{abstract}
  Existing ML-based atmospheric models are not suitable for climate prediction, which requires long-term stability and physical consistency. We present ACE (\AI Climate Emulator), a 200M-parameter, autoregressive machine learning emulator of an existing comprehensive 100-km resolution global atmospheric model. The formulation of ACE allows evaluation of physical laws such as the conservation of mass and moisture. The emulator is stable for 100 years, nearly conserves column moisture without explicit constraints and faithfully reproduces the reference model's climate, outperforming a challenging baseline on over 90\% of tracked variables. ACE requires nearly 100x less wall clock time and is 100x more energy efficient than the reference model using typically available resources. Without fine-tuning, ACE can stably generalize to a previously unseen historical sea surface temperature dataset.
\end{abstract}

\section{Introduction}

The last year has seen a revolution in the field of numerical weather prediction. Multiple groups have shown improvements in key metrics over the state of the art physics-based medium-range weather prediction system using deep learning methods \cite{PanguWeather,FengWu,GraphCast}. However, the applicability of these methods to climate modeling is unclear. Nearly all machine learning based weather prediction systems \cite{PanguWeather, FengWu, Keisler2022, GraphCast, FourCastNet} have reported results for forecasts up to only 14 days---with notable exceptions of \cite{Bonev2023,Weyn2020}---and instabilities or unphysical artifacts often occur for longer simulations.

We claim the requirements of an ML-based atmospheric model for climate prediction are as follows. Such a model should maintain realistic weather variability and be stable for indefinite periods. Conservation of mass, moisture and energy is key. Surface and top-of-atmosphere fluxes of energy, moisture and momentum must be predicted to enable assessment of climate sensitivity and coupling with other components such as the ocean. Appropriate forcings should be used---e.g.\ sea surface temperature (SST) in the case of an atmosphere-only model. Its long-term averages should be unbiased compared to a reference dataset. Finally, the model's performance must generalize across a broad range of plausible SST distributions and CO$_2$ concentrations.

Here we present ACE (\AI Climate Emulator)\footnote{Code, data and model weights are publicly available; see https://github.com/ai2cm/ace.}, a neural network based atmospheric model which satisfies many of the criteria listed above. ACE uses the Spherical Fourier Neural Operator (SFNO) architecture \cite{Bonev2023} and is trained to emulate an existing physics-based atmospheric model with 6-hour temporal resolution. ACE runs stably for at least 100 years and can simulate a decade in one hour of wall clock time, nearly 100 times faster than the reference atmospheric model, while consuming 100 times less energy. ACE predicts diagnostics such as the fluxes of energy and moisture through the top of atmosphere and Earth surface (e.g.\ precipitation). The model is framed so that precise evaluation of conservation of mass and moisture is possible and we find that column moisture is very nearly conserved across individual timesteps. External forcings, such as incoming solar radiation and sea surface temperature, are used as inputs. Finally, ACE replicates the near-surface climatology of the reference model better and runs much faster than a 2x coarser but otherwise identical version of that model.

Related work includes ClimateBench \cite{WatsonParris2022} which proposes directly predicting climate metrics such as annual mean precipitation from input forcing variables like CO$_2$. The disadvantage of such an approach is the limited physical interpretability: for example what sub-annual variability gives rise to the annual mean? Another study \cite{ClimaX} trains on climate model output but makes forecasts with 14- or 30-day lead times, leading to smooth predictions near the climatological mean (e.g.\ Fig.\ 20 of \cite{ClimaX}).

\section{Methods}

\subsection{Dataset}
\looseness=-1%
Most ML-based weather prediction systems have been trained on the ERA5 reanalysis dataset \cite{Hersbach2020}. While appealing due to its relatively accurate representation of historical atmospheric conditions, reanalysis data has downsides for the development of machine learning based climate models: it has a limited number of samples restricted to the recent past; models trained on reanalysis may not be reliable for future climates \cite{Beucler2021,Clark2022}; and analysis increment terms (adjustments from observations) have no clear physical interpretation \cite{Trenberth2011}. Therefore, we generate training data with an existing global atmospheric model (FV3GFS, the atmospheric component of the United States weather model \cite{UFS_2020,Zhou2019}).

The training data are an 11-member initial condition ensemble of 10-year simulations (hereafter the ``reference'' simulation; 10 years is length after discarding 3-month spinup time) performed on NOAA/GFDL's GAEA computer. Ten ensemble members are used for training and the eleventh for validation. For simplicity, we use annually repeating climatological sea surface temperature (1982-2012 average) and fixed greenhouse gas and aerosol concentrations. The reference simulation has a cubed-sphere grid \cite{Putman2007}  with a horizontal spacing of about 100$\,$km and 63 vertical layers. Model state is saved every 6 hours, with a combination of snapshot and interval-mean variables. See Table~\ref{table:variables} for a complete description of variables used for training. For compatibility with SFNO we regrid conservatively from the cubed-sphere geometry of FV3GFS to a~1° Gaussian grid \cite{Washington2005}, additionally filtering the data with a spherical harmonic transform round-trip to remove artifacts in the high latitudes. We coarsen the vertical coordinate to~8 layers while conserving moisture and energy (Appendix~\ref{appendix:data}).

\subsection{Training}
\label{subsec:training}
\paragraph{Architecture}
\looseness=-1%
We use the SFNO architecture \cite{Bonev2023} to predict the state of the atmosphere at time $t+6\mathrm{hr}$ using the state at time $t$ as input. SFNO is a Fourier Neural Operator-based architecture which uses spherical harmonic transforms to enable efficient global convolutions on the sphere, while respecting inherent symmetries of the spherical domain. Hyperparameters are described in Appendix~\ref{appendix:hyperparameters}; the number of parameters is about 200M. Unlike many prior ML atmospheric prediction systems, we use prognostic variables $P$ which are both inputs and outputs, forcing variables $F$ which are inputs only and diagnostic variables $D$ which are outputs only (Table~\ref{table:variables}). Explicitly, with $t$ representing the time index: $[P_{t+1}, D_{t+1}] = f(P_t, F_t)$, where $f$ represents the SFNO module and forcing variables $F_t$ are read from an external dataset (Figure~\ref{fig:variable_flow_diagram}).
Variables are chosen to be forcing, prognostic or diagnostic based on how they are used in the reference physics-based simulation. The diagnostic variables do not inform the next step, which is typical for physics-based atmospheric models and has the important benefit that fields such as precipitation are not necessary to initialize a simulation. However, because they are predicted by the same architecture that predicts the prognostic variables one can enforce physical constraints such as moisture conservation which affect prognostic model weights. This is different than some previous approaches which use a separate model to predict precipitation \cite{FourCastNet}. 

\paragraph{Data Normalization}
Variables are normalized using a ``residual scaling'' approach such that predicting outputs equal to input would result in each variable contributing equally to the loss function (similar to \cite{Keisler2022}). See Appendix~\ref{appendix:normalization} for details and Appendix~\ref{appendix:normalizationablation} for an ablation of this choice. This ``residual scaling'' approach has the largest impact on the surface pressure, which ends up having a normalized standard deviation about 20 times larger than otherwise (Figure~\ref{fig:normstddev}).

\paragraph{Loss Function}
Given $\mathbf{x}_i$ representing the normalized target for the $i$'th sample of a batch for all spatial points and channels, and $\mathbf{\hat{x}}_i$ as the corresponding prediction, the loss for a batch of size $B$ is $\frac{1}{B} \sum_{i=1}^{B} \frac{\lVert \mathbf{\hat{x}}_i - \mathbf{x}_i \rVert_2}{\lVert \mathbf{x}_i \rVert_2}$. The loss is computed after a single 6-hour forward step. Optimization hyperparameters are listed in Table~\ref{table:optimizationparams}.

\subsection{Evaluation}
We are not aware of any existing purely machine learning based system that allows for long (at least 10-year) forecasts and includes a vertically resolved view of the entire atmosphere. Therefore, we formulate baselines using our physics-based model. Two cases will be considered: first a ``reference" perfect emulator in which we compare members of the initial condition ensemble against each other, giving an upper bound on model skill. Because of the chaotic nature of the atmosphere and the limited duration (10 years) of our validation dataset, even a perfect emulator will have non-zero errors. Second, we run a difficult-to-beat 200$\,$km ``baseline'': the same physics-based FV3GFS model but using both horizontal resolution and dynamics time step that are 2x coarser. This mimics the typical climate modeling strategy for faster simulation: use coarser resolution at the cost of accuracy.

Three metrics are used for evaluation: the time-dependent area-weighted global mean and the area-weighted global mean bias and RMSE of time-mean fields. These are defined in Appendix~\ref{appendix:metrics}.

\subsection{Random seed ensemble}
Because long-term climate skill is not entirely constrained by the 6-hour prediction that is our training objective, we train an ensemble of 5 models which vary only in terms of their initialization random seed. Although validation loss is very similar between the models and all models are capable of stable 10-year forecasts, the magnitude of climate biases varies by up to a factor of 4 (depending on output variable). Hereafter, we report results for the model which performs best on the time-mean RMSE metric for $T_7$ and total water path (TWP) computed over a single 10-year forecast (see Figure~\ref{fig:randomseed}).

\section{Results}
\label{sec:results}
\begin{figure}
  \centering
  \includegraphics[width=\textwidth]{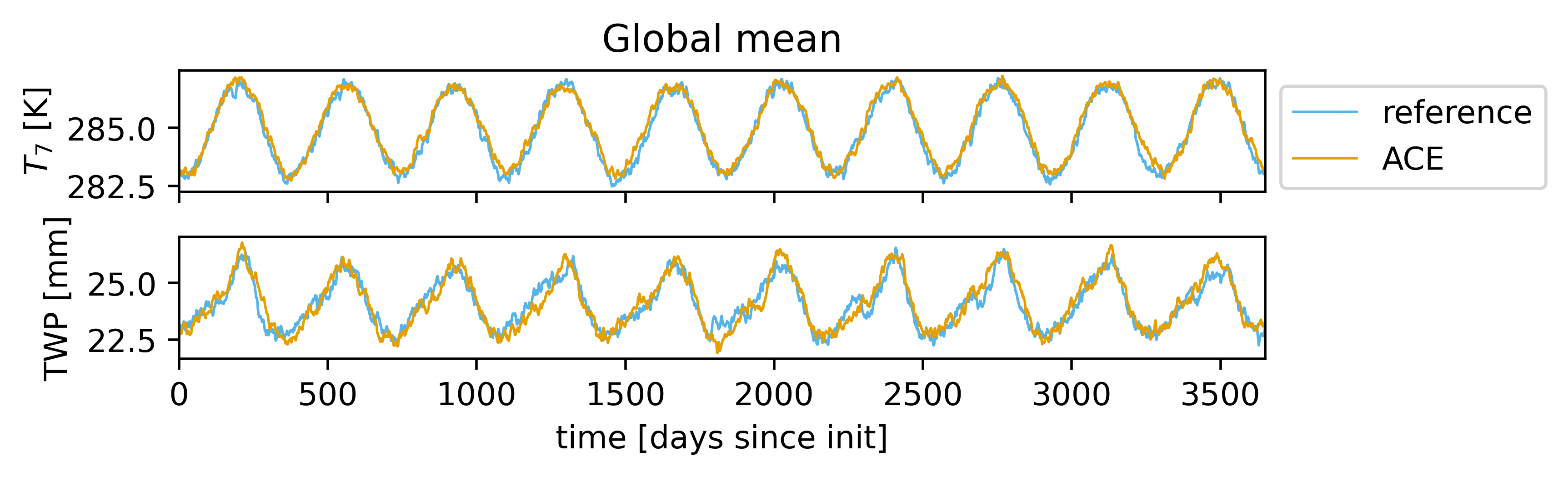}
  \caption{Global mean timeseries of (top) near-surface air temperature $T_7$ and (bottom) total water path computed as $\mathrm{TWP}=\frac{1}{g}\sum_k q_k^T \, dp_k$. For clarity, the daily average is plotted.}
  \label{fig:globaltimeseries}
\end{figure}

\paragraph{Long-term stability}
\looseness=-1%
Initializing from the start of the validation dataset, ACE is able to maintain a stable simulation and an unbiased global mean evolution of temperature and total water path for at least 10 years (Figure~\ref{fig:globaltimeseries}). While there is some year-to-year variability, the seasonal cycle of both of these fields is well represented by ACE. This result is dependent on choice of forcing variables, the normalization strategy (Appendix~\ref{appendix:normalization}) and using the SFNO architecture which has been shown to have favorable stability properties compared to Fourier Transform-based FourCastNet \cite{Bonev2023}. The two prognostic variables whose global means drift unrealistically are the surface pressure $p_s$ and upper-stratospheric water $q^T_0$ (Figure~\ref{fig:globaltimeseriesall}). This run has been extended to 100 years with no sign of instability or long-term drift (Figure~\ref{fig:100yearrun}). However, there are unrealistically large annual variations in certain fields, such as global mean precipitation rate.

\begin{figure}
  \centering
  \includegraphics[width=1\textwidth]{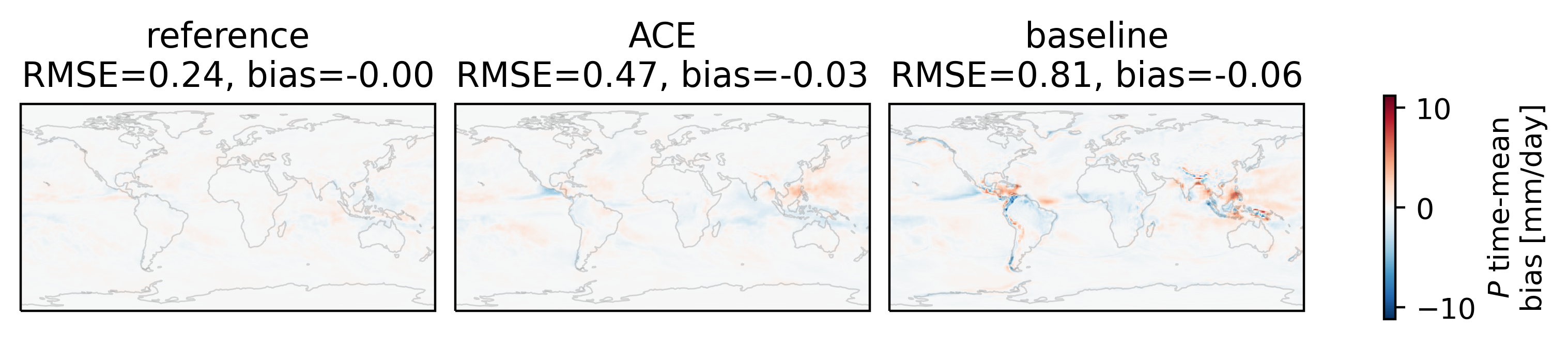}
  \caption{10-year mean bias in surface precipitation rate. Titles show global and time-mean RMSE and bias in units of mm/day (Equations~\ref{eq:globaltimermse} and \ref{eq:globaltimebias}).}
  \label{fig:precipmaps}
\end{figure}

\paragraph{Climate biases}
With long-term stability comes the possibility of quantitatively evaluating climate biases. Figure~\ref{fig:precipmaps} shows the time-averaged bias in the spatial pattern of surface precipitation rate. ACE has impressively small precipitation biases, with a global RMSE about 42\% smaller than the baseline. The time-mean biases and RMSEs are shown for all output channels in Appendix~\ref{appendix:results}. 41 out of 44 output variables have a lower time-mean RMSE for ACE compared to the baseline (Figure~\ref{fig:globaltimermse}).

\begin{figure}
  \centering
  \includegraphics[width=1\textwidth]{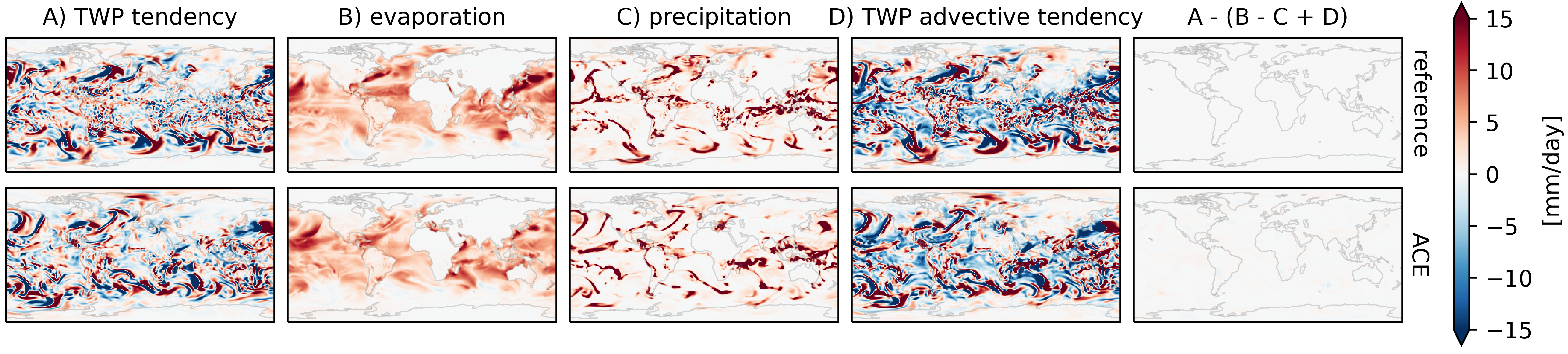}
  \caption{Snapshot of the terms in the column moisture budget (Equation~\ref{eq:twp}) one year into simulation for (top) reference data and (bottom) ACE simulation. Given chaotic nature of atmosphere, we do not expect details to match between the reference and ACE simulations. If column-integrated moisture is exactly conserved, the rightmost column should equal zero, as it is for the reference data.}
  \label{fig:columnmoisture}
\end{figure}

\paragraph{Physical consistency}
Many aspects of physical consistency are important for interpretability in climate model simulations. Here we discuss one example, the conservation of water expressed in terms of total water path:
\begin{equation}
   \label{eq:twp}
   \frac{\partial TWP}{\partial t} = E - P + \left. \frac{\partial TWP}{\partial t}\right |_{adv}
 \end{equation}
where $TWP = \frac{1}{g} \sum_k q^T_k \, dp_k$ is the amount of water in an atmospheric column and $\left. \frac{\partial TWP}{\partial t}\right |_{adv}$ is the tendency of the total water path due to advection. $E$ and $P$ are the surface evaporation ($E=LHF/L_v$) and precipitation rate respectively. The physical model exactly satisfies Equation~\ref{eq:twp} by design but an ML emulator may not unless explicitly designed to do so. Nonetheless, ACE very nearly obeys the column-wise conservation of moisture (Eq.~\ref{eq:twp}). Figure~\ref{fig:columnmoisture} shows the magnitude of the violation of the budget is very small compared to the individual terms in the budget: standard deviation of total water path tendency is $\sim$50 times that of the column budget violation term. This is true even one year into the inference simulation. Global mean budgets are described in Appendix~\ref{appendix:globalbudgets}.

\begin{figure}
  \centering
  \includegraphics[width=1\textwidth]{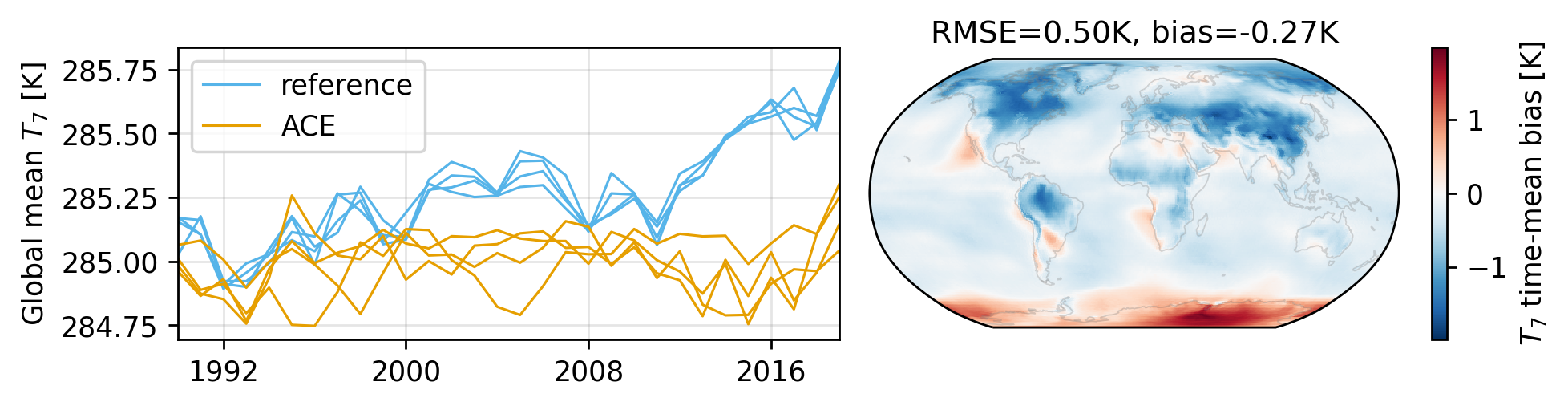}
  \caption{ACE forced with a realistic SST dataset spanning 1990-2020. Showing (left) timeseries of annual mean and global mean $T_7$, with individual lines corresonding to different initial conditions and (right) time-mean bias of $T_7$ for a single initial condition with time-mean RMSE and global bias (Equations ~\ref{eq:globaltimermse} and \ref{eq:globaltimebias}) reported in the title. Here, the ``reference'' is the physics based model (FV3GFS) run for the same period and forced by the same SST pattern.}
  \label{fig:amipSST}
\end{figure}

\paragraph{Zero-shot generalization to realistic sea surface temperature forcing}
In this work, as a first effort, we chose to train on a simplified dataset with annually repeating climatological SST. However, to be useful as a climate model, ACE must be skillful when forced by a wider variety of more realistic SST patterns. In this section we show results of zero-shot generalization to a forcing dataset with historical SSTs from 1990 to 2020 (using CMIP forcing data \cite{Eyring2016}). ACE is stable over the full 30-year span of this previously unseen forcing dataset (Figure~\ref{fig:amipSST}a). This is particularly impressive because the climatological SST used to generate training data spans 1982-2012 and so the realistic SST dataset includes temperatures warmer than any in the training dataset. When forced with the realistic SST dataset, ACE has a cold near-surface temperature bias which is particularly amplified over land where the surface temperatures are not directly forced (Figure~\ref{fig:amipSST}b) and it does not reproduce the interannual variability of the reference. The time-mean RMSE and bias of $T_7$ (0.5K and -0.27K, respectively) are somewhat larger than those when comparing ACE to a reference run which uses the climatological SST dataset (0.35K and 0.12K; see Figures~\ref{fig:globaltimermse} and \ref{fig:globaltimebias}). Nevertheless, the stability and modest biases of ACE when using this previously unseen SST dataset as a forcing are reassuring.

\paragraph{Computational expense}
Training time for each seed was about 63 hours on four NVIDIA-A100s. Running inference on a single A100 requires about one second of wall clock time per simulated day. For comparison, the reference simulation ran on 96 cores of AMD EPYC 7H12 processors and took $\sim$77 seconds per simulated day. The 2x coarser resolution baseline ran on 24 cores in $\sim$45 seconds per simulated day.

\section{Conclusions and future work}
\looseness=-1%
This work demonstrates the potential of deep learning for skillful and fast climate model emulation. 100x speed-up in run time and 100x greater energy efficiency could democratize the use of climate models, open new research avenues and potentially reduce energy usage. However, there are additional steps before this system is a useful climate model \cite{Eyring2016}. Generalizability is a key challenge. To reduce confounding factors, in this study we focused on a training dataset with simplified (annually-repeating) forcing. It may be necessary to expand the training and input variable set to be able to handle a changing climate. Our suggested approach is to train on a broad range of simulation data that covers the regimes of interest \cite{Clark2022}. However, simulated data are not the real world and have their share of biases. Potential solutions are to fine-tune on reanalysis data (e.g.\ \cite{Ham2019}) or on a smaller amount of high-resolution simulation data that has smaller biases \cite{Bretherton2022}. Further improvements to our current training regime that are possible, e.g. using  appropriate constraints in the loss function \cite{Raissi2019} to reduce global non-physical sources of moisture and mass. Finally, coupling to other components (ocean, sea-ice, land) of the climate system is necessary. Tackling these challenges is an exciting opportunity for the growing field of machine learning based climate modeling.

\vspace{-1mm}
\begin{ack}
\vspace{-2mm}
We acknowledge NOAA's Geophysical Fluid Dynamics Laboratory for providing the computing resources used to perform the reference FV3GFS simulations. This research used resources of NERSC, a U.S. Department of Energy Office of Science User Facility located at Lawrence Berkeley National Laboratory, using NERSC award BER-ERCAP0024832.
\end{ack}

\bibliography{main}

\begin{thebibliography}{24}
\providecommand{\natexlab}[1]{#1}
\providecommand{\url}[1]{\texttt{#1}}
\expandafter\ifx\csname urlstyle\endcsname\relax
  \providecommand{\doi}[1]{doi: #1}\else
  \providecommand{\doi}{doi: \begingroup \urlstyle{rm}\Url}\fi

\bibitem[Beucler et~al.(2021)Beucler, Gentine, Yuval, Gupta, Peng, Lin, Yu,
  Rasp, Ahmed, O'Gorman, Neelin, Lutsko, and Pritchard]{Beucler2021}
Tom Beucler, Pierre Gentine, Janni Yuval, Ankitesh Gupta, Liran Peng, Jerry
  Lin, Sungduk Yu, Stephan Rasp, Fiaz Ahmed, Paul~A. O'Gorman, J.~David Neelin,
  Nicholas~J. Lutsko, and Michael Pritchard.
\newblock Climate-invariant machine learning.
\newblock \emph{arXiv}, 2021.
\newblock \doi{10.48550/ARXIV.2112.08440}.

\bibitem[Bi et~al.(2023)Bi, Xie, Zhang, Chen, Gu, and Tian]{PanguWeather}
Kaifeng Bi, Lingxi Xie, Hengheng Zhang, Xin Chen, Xiaotao Gu, and Qi~Tian.
\newblock Accurate medium-range global weather forecasting with 3d neural
  networks.
\newblock \emph{Nature}, 619\penalty0 (7970):\penalty0 533--538, July 2023.
\newblock \doi{10.1038/s41586-023-06185-3}.

\bibitem[Bonev et~al.(2023)Bonev, Kurth, Hundt, Pathak, Baust, Kashinath, and
  Anandkumar]{Bonev2023}
Boris Bonev, Thorsten Kurth, Christian Hundt, Jaideep Pathak, Maximilian Baust,
  Karthik Kashinath, and Anima Anandkumar.
\newblock Spherical fourier neural operators: Learning stable dynamics on the
  sphere.
\newblock \emph{Proceedings of the 40th International Conference on Machine
  Learning (ICML)}, 2023.
\newblock \doi{10.48550/ARXIV.2306.03838}.

\bibitem[Bretherton et~al.(2022)Bretherton, Henn, Kwa, Brenowitz, Watt-Meyer,
  McGibbon, Perkins, Clark, and Harris]{Bretherton2022}
Christopher~S. Bretherton, Brian Henn, Anna Kwa, Noah~D. Brenowitz, Oliver
  Watt-Meyer, Jeremy McGibbon, W.~Andre Perkins, Spencer~K. Clark, and Lucas
  Harris.
\newblock Correcting coarse-grid weather and climate models by machine learning
  from global storm-resolving simulations.
\newblock \emph{Journal of Advances in Modeling Earth Systems}, 14\penalty0
  (2), February 2022.
\newblock \doi{10.1029/2021ms002794}.

\bibitem[Chen et~al.(2023)Chen, Han, Gong, Bai, Ling, Luo, Chen, Ma, Zhang, Su,
  Ci, Li, Yang, and Ouyang]{FengWu}
Kang Chen, Tao Han, Junchao Gong, Lei Bai, Fenghua Ling, Jing-Jia Luo, Xi~Chen,
  Leiming Ma, Tianning Zhang, Rui Su, Yuanzheng Ci, Bin Li, Xiaokang Yang, and
  Wanli Ouyang.
\newblock Fengwu: Pushing the skillful global medium-range weather forecast
  beyond 10 days lead.
\newblock \emph{arXiv}, 2023.
\newblock \doi{10.48550/ARXIV.2304.02948}.

\bibitem[Clark et~al.(2022)Clark, Brenowitz, Henn, Kwa, McGibbon, Perkins,
  Watt-Meyer, Bretherton, and Harris]{Clark2022}
Spencer~K. Clark, Noah~D. Brenowitz, Brian Henn, Anna Kwa, Jeremy McGibbon,
  W.~Andre Perkins, Oliver Watt-Meyer, Christopher~S. Bretherton, and Lucas~M.
  Harris.
\newblock Correcting a 200~km resolution climate model in multiple climates by
  machine learning from 25~km resolution simulations.
\newblock \emph{Journal of Advances in Modeling Earth Systems}, 14\penalty0
  (9), September 2022.
\newblock \doi{10.1029/2022ms003219}.

\bibitem[Collins et~al.(2004)Collins, Rasch, Boville, McCaa, Williamson, Kiehl,
  Briegleb, Bitz, Lin, Zhang, and Dai]{Collins2004}
William Collins, Philip Rasch, Byron Boville, James McCaa, David Williamson,
  Jeffrey Kiehl, Bruce Briegleb, Cecilia Bitz, S.-J. Lin, Minghua Zhang, and
  Youngjiu Dai.
\newblock Description of the {NCAR Community Atmosphere Model (CAM 3.0)}.
\newblock Technical report, {UCAR/NCAR}, 2004.

\bibitem[Eyring et~al.(2016)Eyring, Bony, Meehl, Senior, Stevens, Stouffer, and
  Taylor]{Eyring2016}
Veronika Eyring, Sandrine Bony, Gerald~A. Meehl, Catherine~A. Senior, Bjorn
  Stevens, Ronald~J. Stouffer, and Karl~E. Taylor.
\newblock Overview of the coupled model intercomparison project phase 6
  ({CMIP}6) experimental design and organization.
\newblock \emph{Geoscientific Model Development}, 9\penalty0 (5):\penalty0
  1937--1958, May 2016.
\newblock \doi{10.5194/gmd-9-1937-2016}.

\bibitem[Ham et~al.(2019)Ham, Kim, and Luo]{Ham2019}
Yoo-Geun Ham, Jeong-Hwan Kim, and Jing-Jia Luo.
\newblock Deep learning for multi-year {ENSO} forecasts.
\newblock \emph{Nature}, 573\penalty0 (7775):\penalty0 568--572, September
  2019.
\newblock \doi{10.1038/s41586-019-1559-7}.

\bibitem[Hersbach et~al.(2020)Hersbach, Bell, Berrisford, Hirahara,
  Hor{\'{a}}nyi, Mu{\~{n}}oz-Sabater, Nicolas, Peubey, Radu, Schepers, Simmons,
  Soci, Abdalla, Abellan, Balsamo, Bechtold, Biavati, Bidlot, Bonavita, Chiara,
  Dahlgren, Dee, Diamantakis, Dragani, Flemming, Forbes, Fuentes, Geer,
  Haimberger, Healy, Hogan, H{\'{o}}lm, Janiskov{\'{a}}, Keeley, Laloyaux,
  Lopez, Lupu, Radnoti, Rosnay, Rozum, Vamborg, Villaume, and
  Th{\'{e}}paut]{Hersbach2020}
Hans Hersbach, Bill Bell, Paul Berrisford, Shoji Hirahara, Andr{\'{a}}s
  Hor{\'{a}}nyi, Joaqu{\'{\i}}n Mu{\~{n}}oz-Sabater, Julien Nicolas, Carole
  Peubey, Raluca Radu, Dinand Schepers, Adrian Simmons, Cornel Soci, Saleh
  Abdalla, Xavier Abellan, Gianpaolo Balsamo, Peter Bechtold, Gionata Biavati,
  Jean Bidlot, Massimo Bonavita, Giovanna Chiara, Per Dahlgren, Dick Dee,
  Michail Diamantakis, Rossana Dragani, Johannes Flemming, Richard Forbes,
  Manuel Fuentes, Alan Geer, Leo Haimberger, Sean Healy, Robin~J. Hogan,
  El{\'{\i}}as H{\'{o}}lm, Marta Janiskov{\'{a}}, Sarah Keeley, Patrick
  Laloyaux, Philippe Lopez, Cristina Lupu, Gabor Radnoti, Patricia Rosnay,
  Iryna Rozum, Freja Vamborg, Sebastien Villaume, and Jean-Noël Th{\'{e}}paut.
\newblock The {ERA}5 global reanalysis.
\newblock \emph{Quarterly Journal of the Royal Meteorological Society},
  146\penalty0 (730):\penalty0 1999--2049, June 2020.
\newblock \doi{10.1002/qj.3803}.

\bibitem[Keisler(2022)]{Keisler2022}
Ryan Keisler.
\newblock Forecasting global weather with graph neural networks.
\newblock \emph{arXiv}, 2022.
\newblock \doi{10.48550/ARXIV.2202.07575}.

\bibitem[Kucharski et~al.(2013)Kucharski, Molteni, King, Farneti, Kang, and
  Feudale]{Kucharski2013}
Fred Kucharski, Franco Molteni, Martin~P. King, Riccardo Farneti, In-Sik Kang,
  and Laura Feudale.
\newblock On the need of intermediate complexity general circulation models: A
  {\textquotedblleft}{SPEEDY}{\textquotedblright} example.
\newblock \emph{Bulletin of the American Meteorological Society}, 94\penalty0
  (1):\penalty0 25--30, January 2013.
\newblock \doi{10.1175/bams-d-11-00238.1}.

\bibitem[Lam et~al.(2022)Lam, Sanchez-Gonzalez, Willson, Wirnsberger,
  Fortunato, Alet, Ravuri, Ewalds, Eaton-Rosen, Hu, Merose, Hoyer, Holland,
  Vinyals, Stott, Pritzel, Mohamed, and Battaglia]{GraphCast}
Remi Lam, Alvaro Sanchez-Gonzalez, Matthew Willson, Peter Wirnsberger, Meire
  Fortunato, Ferran Alet, Suman Ravuri, Timo Ewalds, Zach Eaton-Rosen, Weihua
  Hu, Alexander Merose, Stephan Hoyer, George Holland, Oriol Vinyals, Jacklynn
  Stott, Alexander Pritzel, Shakir Mohamed, and Peter Battaglia.
\newblock Graphcast: Learning skillful medium-range global weather forecasting.
\newblock \emph{arXiv}, 2022.
\newblock \doi{10.48550/ARXIV.2212.12794}.

\bibitem[Li et~al.(2022)Li, Huang, Liu, and Anandkumar]{li2022geofno}
Zongyi Li, Daniel~Zhengyu Huang, Burigede Liu, and Anima Anandkumar.
\newblock Fourier neural operator with learned deformations for pdes on general
  geometries.
\newblock \emph{arXiv}, 2022.
\newblock \doi{10.48550/arXiv.2207.05209}.

\bibitem[Nguyen et~al.(2023)Nguyen, Brandstetter, Kapoor, Gupta, and
  Grover]{ClimaX}
Tung Nguyen, Johannes Brandstetter, Ashish Kapoor, Jayesh~K. Gupta, and Aditya
  Grover.
\newblock Climax: A foundation model for weather and climate.
\newblock \emph{arXiv}, 2023.
\newblock \doi{10.48550/ARXIV.2301.10343}.

\bibitem[Pathak et~al.(2022)Pathak, Subramanian, Harrington, Raja,
  Chattopadhyay, Mardani, Kurth, Hall, Li, Azizzadenesheli, Hassanzadeh,
  Kashinath, and Anandkumar]{FourCastNet}
Jaideep Pathak, Shashank Subramanian, Peter Harrington, Sanjeev Raja, Ashesh
  Chattopadhyay, Morteza Mardani, Thorsten Kurth, David Hall, Zongyi Li, Kamyar
  Azizzadenesheli, Pedram Hassanzadeh, Karthik Kashinath, and Animashree
  Anandkumar.
\newblock Fourcastnet: A global data-driven high-resolution weather model using
  adaptive fourier neural operators.
\newblock \emph{arXiv}, 2022.
\newblock \doi{10.48550/ARXIV.2202.11214}.

\bibitem[Putman and Lin(2007)]{Putman2007}
William~M. Putman and Shian-Jiann Lin.
\newblock Finite-volume transport on various cubed-sphere grids.
\newblock \emph{Journal of Computational Physics}, 227\penalty0 (1):\penalty0
  55--78, November 2007.
\newblock \doi{10.1016/j.jcp.2007.07.022}.

\bibitem[Raissi et~al.(2019)Raissi, Perdikaris, and Karniadakis]{Raissi2019}
M.~Raissi, P.~Perdikaris, and G.E. Karniadakis.
\newblock Physics-informed neural networks: A deep learning framework for
  solving forward and inverse problems involving nonlinear partial differential
  equations.
\newblock \emph{Journal of Computational Physics}, 378:\penalty0 686--707,
  February 2019.
\newblock \doi{10.1016/j.jcp.2018.10.045}.

\bibitem[Trenberth et~al.(2011)Trenberth, Fasullo, and Mackaro]{Trenberth2011}
Kevin~E. Trenberth, John~T. Fasullo, and Jessica Mackaro.
\newblock Atmospheric moisture transports from ocean to land and global energy
  flows in reanalyses.
\newblock \emph{Journal of Climate}, 24\penalty0 (18):\penalty0 4907--4924,
  September 2011.
\newblock \doi{10.1175/2011jcli4171.1}.

\bibitem[{UFS Community}(2020)]{UFS_2020}
{UFS Community}.
\newblock Unified forecast system {(UFS)}.
\newblock \emph{UFS}, Oct 2020.
\newblock \doi{10.5281/zenodo.4460292}.

\bibitem[Washington and Parkinson(2005)]{Washington2005}
Warren~M. Washington and Claire~L. Parkinson.
\newblock \emph{An Introduction to Three-Dimensional Climate Modeling}.
\newblock University Science Books, Sausalito, CA, 2005.

\bibitem[Watson-Parris et~al.(2022)Watson-Parris, Rao, Olivi{\'{e}}, Seland,
  Nowack, Camps-Valls, Stier, Bouabid, Dewey, Fons, Gonzalez, Harder, Jeggle,
  Lenhardt, Manshausen, Novitasari, Ricard, and Roesch]{WatsonParris2022}
D.~Watson-Parris, Y.~Rao, D.~Olivi{\'{e}}, {\O}.~Seland, P.~Nowack,
  G.~Camps-Valls, P.~Stier, S.~Bouabid, M.~Dewey, E.~Fons, J.~Gonzalez,
  P.~Harder, K.~Jeggle, J.~Lenhardt, P.~Manshausen, M.~Novitasari, L.~Ricard,
  and C.~Roesch.
\newblock {ClimateBench} v1.0: A benchmark for data-driven climate projections.
\newblock \emph{Journal of Advances in Modeling Earth Systems}, 14\penalty0
  (10), October 2022.
\newblock \doi{10.1029/2021ms002954}.

\bibitem[Weyn et~al.(2020)Weyn, Durran, and Caruana]{Weyn2020}
Jonathan~A. Weyn, Dale~R. Durran, and Rich Caruana.
\newblock Improving data-driven global weather prediction using deep
  convolutional neural networks on a cubed sphere.
\newblock \emph{Journal of Advances in Modeling Earth Systems}, 12\penalty0
  (9), September 2020.
\newblock \doi{10.1029/2020ms002109}.

\bibitem[Zhou et~al.(2019)Zhou, Lin, Chen, Harris, Chen, and Rees]{Zhou2019}
Linjiong Zhou, Shian-Jiann Lin, Jan-Huey Chen, Lucas~M. Harris, Xi~Chen, and
  Shannon~L. Rees.
\newblock Toward convective-scale prediction within the next generation global
  prediction system.
\newblock \emph{Bulletin of the American Meteorological Society}, 100\penalty0
  (7):\penalty0 1225 -- 1243, 2019.
\newblock \doi{https://doi.org/10.1175/BAMS-D-17-0246.1}.

\end{thebibliography}
\bibliographystyle{plainnat}

\newpage
\appendix

\section{Data preprocessing}
\label{appendix:data}

\begin{figure}

  \centering
\includegraphics[width=\textwidth]{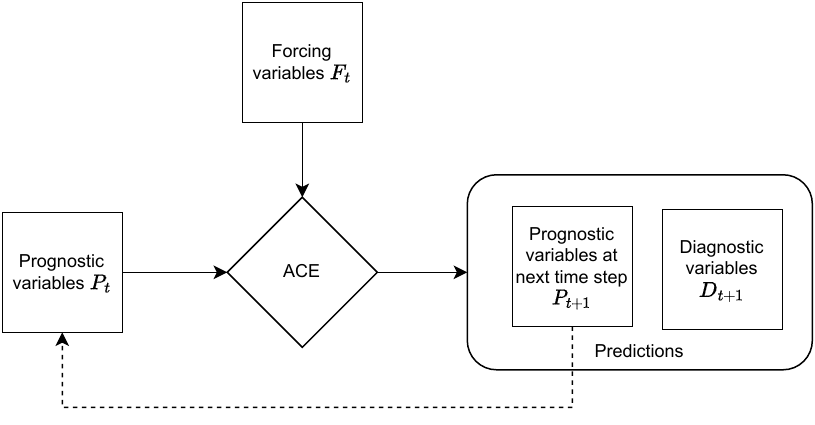}
\caption{Diagram summarizing the flow of input and output variables. Prognostic variables are fed back into the model autoregressively. Forcing variables are read from an external dataset and appended to the prognostic variables at each step. The network outputs diagnostic variables, which contribute to the loss but are not passed back as inputs for the next step.}\label{fig:variable_flow_diagram}
\end{figure}

The complete list of input and output variables used for ACE is given in Table~\ref{table:variables}. A schematic depiction of the meaning of forcing, prognostic and diagnostic variables is shown in Figure~\ref{fig:variable_flow_diagram}.

\begin{table}
  \caption{Input and output variables for ACE. The $k$ subscript refers to a vertical layer index, and ranges from 0 to 7 starting at the top of atmosphere and increasing towards the surface. The Time column indicates whether a variable represents the value at a particular time step (``Snapshot''), the average across the 6-hour time step (``Mean'') or a quantity which does not depend on time (``Invariant''). ``TOA'' denotes ``Top Of Atmosphere'', the climate model's upper boundary.}
  \label{table:variables}
  \centering
  \begin{tabular}{llll}
    \toprule
    \multicolumn{4}{c}{Prognostic (input and output)}   \\
    \midrule
    Symbol   & Description                                & Units & Time        \\
    \midrule
    $T_k$    & Air temperature                            & K     & Snapshot \\
    $q^T_k$   & Specific total water (vapor + condensates) & kg/kg & Snapshot \\
    $u_k$    & Windspeed in eastward direction            & m/s   & Snapshot \\
    $v_k$    & Windspeed in northward direction           & m/s   & Snapshot \\
    $T_s$     & Skin temperature of land or sea-ice            & K     & Snapshot \\
    $p_s$     & Atmospheric pressure at surface            & Pa    & Snapshot \\
    \midrule
    \multicolumn{4}{c}{Forcing (input only)}   \\
    \midrule
    Symbol       & Description                              & Units   & Time          \\
    \midrule
    $DSWRFtoa$ & Downward shortwave radiative flux at TOA & W/m$^2$ & Mean  \\
    $T_s$       & Skin temperature of open ocean           & K      & Snapshot \\
    $z_s$       & Surface height of topography             & m      & Invariant \\
    $f_l$       & Land grid cell fraction                  & $-$    & Invariant \\
    $f_o$       & Ocean grid cell fraction                 & $-$    & Snapshot \\
    $f_{si}$    & Sea-ice grid cell fraction               & $-$    & Snapshot \\
    \midrule
    \multicolumn{4}{c}{Diagnostic (output only)}   \\
    \midrule
    Symbol       & Description                                    & Units   & Time    \\
    \midrule
    $USWRFtoa$ & Upward shortwave radiative flux at TOA         & W/m$^2$ & Mean \\
    $ULWRFtoa$ & Upward longwave radiative flux at TOA          & W/m$^2$ & Mean  \\
    $USWRFsfc$ & Upward shortwave radiative flux at surface     & W/m$^2$ & Mean  \\
    $ULWRFsfc$ & Upward longwave radiative flux at surface      & W/m$^2$ & Mean  \\
    $DSWRFsfc$ & Downward shortwave radiative flux at surface   & W/m$^2$ & Mean  \\
    $DLWRFsfc$ & Downward longwave radiative flux at surface    & W/m$^2$ & Mean  \\
    $P$ & Surface precipitation rate (all phases)               & kg/m$^2$/s & Mean  \\
    $\left. \frac{\partial TWP}{\partial t}\right |_{adv}$ & Tendency of total water path from advection & kg/m$^2$/s & Mean \\
    $LHF$ & Surface latent heat flux                            & W/m$^2$ & Mean  \\
    $SHF$ & Surface sensible heat flux                          & W/m$^2$ & Mean  \\
    \bottomrule
  \end{tabular}
\end{table}

\paragraph{Horizontal regridding}
FV3GFS, the model used to generate reference data, uses a cubed sphere grid \cite{Putman2007}. This grid prevents direct use of the SFNO architecture, which in the formulation of \cite{Bonev2023} requires data on a latitude-longitude grid. Therefore, we regrid to latitude-longitude using the ``fregrid'' tool provided by NOAA GFDL. We use the first-order conservative regridding option. This regridding procedure leads to unphysical sharp gradients in the high latitudes for some fields. Therefore as part of the regridding process, we also perform a round-trip with the spherical harmonic transform which has the effect of filtering out these unrealistic artifacts. For the 2x coarser FV3GFS baseline, regridding to the 100km latitude-longitude grid upsampled the dataset from its native resolution and produced more profound artifacts, and so when performing the spherical harmonic transform round-trip on this dataset we truncated the highest-frequency modes (35\% of all modes), which had the effect of reducing the baseline's errors against the reference dataset.

In future work, it would be natural to directly train on the cubed sphere grid data, which would prevent complications associated with regridding. The cubed sphere grid has been used for ML-based weather prediction architectures \cite{Weyn2020} and adaptations of Fourier Neural Operators for arbitrary geometric discretizations have been proven in low dimensional settings that could be readily extended to SFNO~\cite{li2022geofno}.

\paragraph{Vertical coarse-graining}
After the horizontal regridding is done for all variables, anticipating the requirements for precise mass and energy conservation, we vertically integrate the 3D variables (air temperature, specific total water, eastward wind and northward wind) from the original 63 vertical layers to 8 layers. The reference model FV3GFS uses a hybrid sigma-pressure vertical coordinate (e.g.\ Figure 3.1 of \cite{Collins2004}) where the pressure at vertical layer interface $k$ is defined by:
\begin{equation}
\label{eq:akbk}
    p_k = a_k + b_k p_s
\end{equation}
where $a_k$ and $b_k$ are constant coefficients and $p_s$ is the surface pressure. We also use a hybrid sigma-pressure vertical coordinate for ACE, but subselect the $a_k$ and $b_k$ coordinates from FV3GFS to nine interfaces (Table~\ref{table:verticalcoord}) corresponding to eight finite volume vertical layers. The nine interfaces were chosen to have a vertical coordinate that corresponds closely to the one in the SPEEDY model \cite{Kucharski2013}. The uppermost two layers correspond to the stratosphere while the lowermost layer is roughly a boundary layer average. The following equation, using temperature $T$ as an example, is used to transform all 3D variables from the 63-layer to 8-layer vertical coordinate:
\begin{equation}
    \overline{T}_k = \frac{1}{\overline{dp}_k}\sum_{i=I_k}^{I_{k+1}} T_i \, dp_i
\end{equation}
where
\begin{equation}
    \overline{dp}_k = \sum_{i=I_k}^{I_{k+1}}dp_i
\end{equation}
and $dp_i$ and $T_i$ are the pressure thickness and temperature of the $i$'th atmospheric layer in the original vertical coordinate and $I_k$ is the index of the original vertical coordinate (fourth column of Table~\ref{table:verticalcoord}). The pressure thickness $dp_i=p_{i+1} - p_i$ is computed by taking the difference between the vertical interface pressures given by Equation~\ref{eq:akbk}. The use of this finite volume vertical coordinate ensures that exact vertical integrals of quantities such as mass, moisture and energy can be computed and expected to balance with fluxes into and out of the atmosphere.

\begin{table}
  \caption{ACE vertical coordinate. Here $k$ indicates the vertical layer interface ranging from the top of the model's atmosphere $k=0$ to the Earth surface $k=8$. $a_k$ and $b_k$ define the vertical coordinate (Equation~\ref{eq:akbk}) while $I_k$ indicates what the corresponding vertical index is in the 63-layer reference FV3GFS simulation. $p^{ref}_k$ is the pressure at model layer interfaces assuming $p_s=1000$hPa.}
  \label{table:verticalcoord}
  \centering
  \begin{tabular}{lllll}
    \toprule
    $k$     & $a_k$ [Pa]  & $b_k$ [unitless] & $I_k$ & $p^{ref}_k$ [hPa]\\
    \midrule
    0       & 64.247      & 0.0              & 0     & 0.642 \\
    1       & 5167.14603  & 0.0              & 18    & 51.7 \\
    2       & 12905.42546 & 0.01755          & 26    & 147 \\
    3       & 13982.4677  & 0.11746          & 31    & 257 \\
    4       & 12165.28766 & 0.2896           & 36    & 411 \\
    5       & 8910.07678  & 0.49806          & 41    & 587 \\
    6       & 4955.72632  & 0.72625          & 47    & 776 \\
    7       & 2155.78385  & 0.88192          & 53    & 903 \\
    8       & 0.0         & 1.0              & 63    & 1000 \\
    \bottomrule
  \end{tabular}
\end{table}

\section{Hyperparameters}
\label{appendix:hyperparameters}
Table~\ref{table:sfnoparams} lists the SFNO hyperparameters used in this study. See \cite{Bonev2023} for details about the meaning of these parameters. The only modification to the architecture of SFNO made in this work is in the first spherical harmonic transform and the last inverse spherical harmonic transform, where Gauss-Legendre quadrature is used, as our data is on the Gaussian grid as opposed to the equiangular latitude-longitude grid used in \cite{Bonev2023} (see horizontal regridding section of Appendix~\ref{appendix:data}).

Table~\ref{table:optimizationparams} lists the hyperparameters used for optimization. Model parameters were averaged across training step using an exponential moving average (EMA), which led to moderate (up to 15\%) improvements in the time-mean RMSE metric.

\begin{table}
  \caption{SFNO hyperparameters. Names correspond to the definition of the SphericalFourierNeuralOperatorNet class found here: \url{https://github.com/ai2cm/modulus/blob/94f62e1ce2083640829ec12d80b00619c40a47f8/modulus/models/sfno/sfnonet.py\#L292}. All configuration options not listed here are set to the defaults at the linked code.}
  \label{table:sfnoparams}
  \centering
  \begin{tabular}{ll}
    \toprule
    Name     & Value \\
    \midrule
    \texttt{embed\char`_dim}        & 256       \\
    \texttt{filter\char`_type}      & linear    \\
    \texttt{num\char`_layers}       & 8         \\
    \texttt{operator\char`_type}    & dhconv    \\
    \texttt{scale\char`_factor}     & 1         \\
    \texttt{spectral\char`_layers}  & 3         \\
    \bottomrule
  \end{tabular}
\end{table}

\begin{table}
  \caption{Optimization hyperparameters.}
  \label{table:optimizationparams}
  \centering
  \begin{tabular}{ll}
    \toprule
    Name     & Value \\
    \midrule
    Optimizer                        & Adam       \\
    Initial learning rate            & $1 \times 10^{-4}$    \\
    Learning rate schedule           & Cosine annealing (single cycle)        \\
    Number of epochs                 & 30         \\
    Batch size                       & 4          \\
    Exponential moving average decay rate             & 0.9999 \\
    \bottomrule
  \end{tabular}
\end{table}

\section{Metrics}
\label{appendix:metrics}
Suppose we have an arbitrary predicted field $\hat{x}(t, \phi, \lambda)$ which depends on time $t$, latitude $\phi$ and longitude $\lambda$ and its corresponding target/ground truth $x(t, \phi, \lambda)$. Let $A(\phi, \lambda)$ be a normalized weight proportional to the area of a grid cell. Then the three metrics of interest are:
\begin{align}
    \label{eq:globalmean}
    GM_{\hat{x}}(t) &= \sum_{\phi, \lambda} A(\phi, \lambda) \hat{x}(t, \phi, \lambda) \\
    \label{eq:globaltimermse}
    RMSE\textnormal{-}TM_{\hat{x}} &= \sqrt{ \sum_{\phi, \lambda} A(\phi, \lambda) \left [ \underset{t}{\mathrm{mean}} \left \{ \hat{x}(t, \phi, \lambda) -  x(t, \phi, \lambda) \right \} \right ]^2 } \\
    \label{eq:globaltimebias}
    GM\textnormal{-}TM\textnormal{-}BIAS_{\hat{x}} &= \underset{t}{\mathrm{mean}} \, \left [ GM_{\hat{x}}(t) - GM_x(t) \right ]
\end{align}

\section{Random seed ensemble}
\begin{figure}
  \centering
  \includegraphics[width=0.6\textwidth]{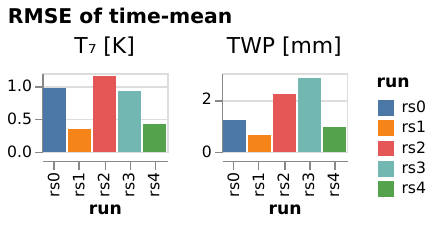}
  \caption{Time-mean RMSE (Equation~\ref{eq:globaltimermse}) of $T_7$ and $\mathrm{TWP}=\frac{1}{g}\sum_k q_k^T \, dp_k$ computed over a single 10-year inference for five ML models which differ only in initialization random seed.}
  \label{fig:randomseed}
\end{figure}

Figure~\ref{fig:randomseed} shows the time-mean RMSE across a 5-member random seed ensemble of models. Although all models are capable of stable 10-year inference runs, the magnitude of the time-mean RMSE metric can vary by up to a factor 4 (depending on output variable being considered). The results in the main body of the paper are all based on the ``rs1'' model which has lowest errors for $T_7$ and total water path. We have verified that the spread in error across different initial conditions is significantly smaller than the spread across random seeds for a single initial condition (not shown).

\section{Systematic evaluation of all outputs}
\label{appendix:results}
\begin{figure}
  \centering
  \includegraphics[width=\textwidth]{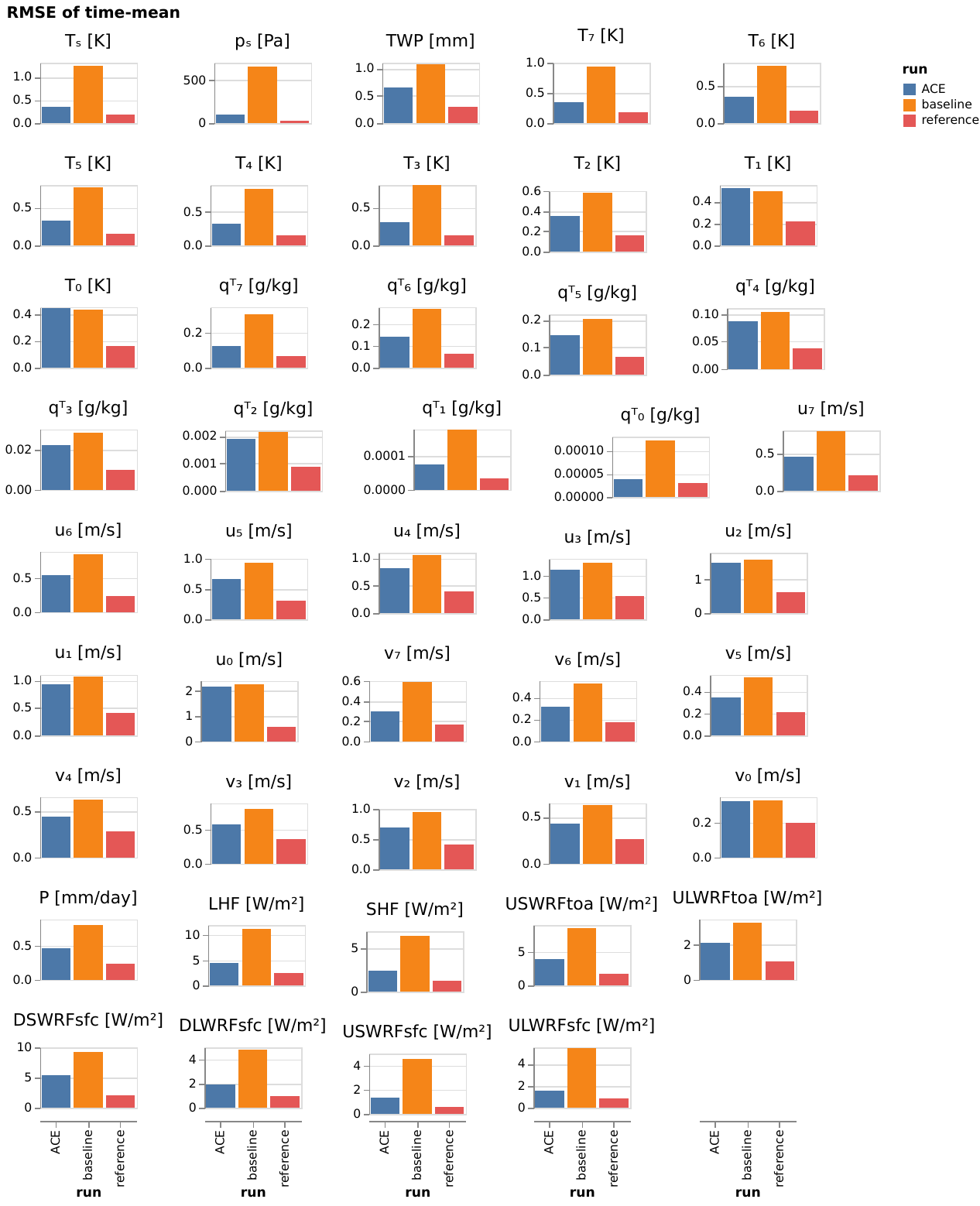}
  \caption{The RMSE of the time-mean (Equation~\ref{eq:globaltimermse}) of all output variables for ACE, the coarser resolution baseline and the reference simulation.}
  \label{fig:globaltimermse}
\end{figure}

In this section we provide a complete description of the global climate biases of ACE relative to the baseline and reference. Figure~\ref{fig:globaltimermse} shows the magnitude of the time-mean pattern errors (Equation~\ref{eq:globaltimermse}) for each output variable except the total water path advective tendency which was not available for the baseline run. Out of the 44 output variables, an impressive 41 have lower RMSE for ACE when compared to the baseline. 

\begin{figure}
  \centering
  \includegraphics[width=0.99\textwidth]{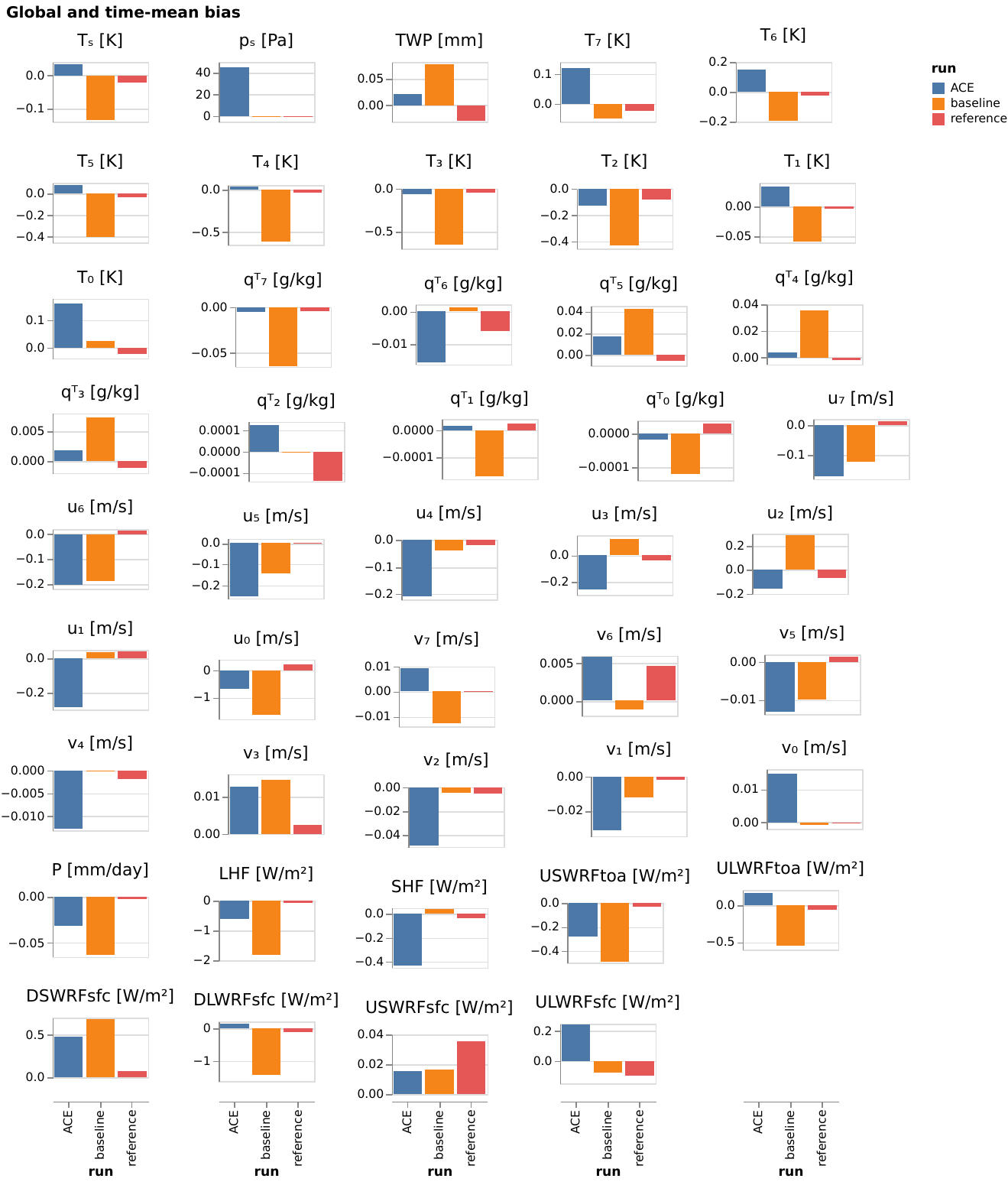}
  \caption{The global- and time-mean bias (Equation~\ref{eq:globaltimebias}) of all output variables for ACE, the coarser resolution baseline and the reference simulation.}
  \label{fig:globaltimebias}
\end{figure}

Figure~\ref{fig:globaltimebias} shows the time- and global-mean bias (Equation~\ref{eq:globaltimebias}) for all output channels. In general biases are low for all variables (e.g.\ radiative flux biases are at most 1.2$\,$W/m$^2$) but they are mostly larger compared to the baseline simulation. In particular the approximately -25 Pa $p_s$ bias is large compared to the near-zero bias in the baseline simulation which exactly conserves global dry air mass and moisture.

\begin{figure}
  \centering
  \includegraphics[width=1.1\textwidth]{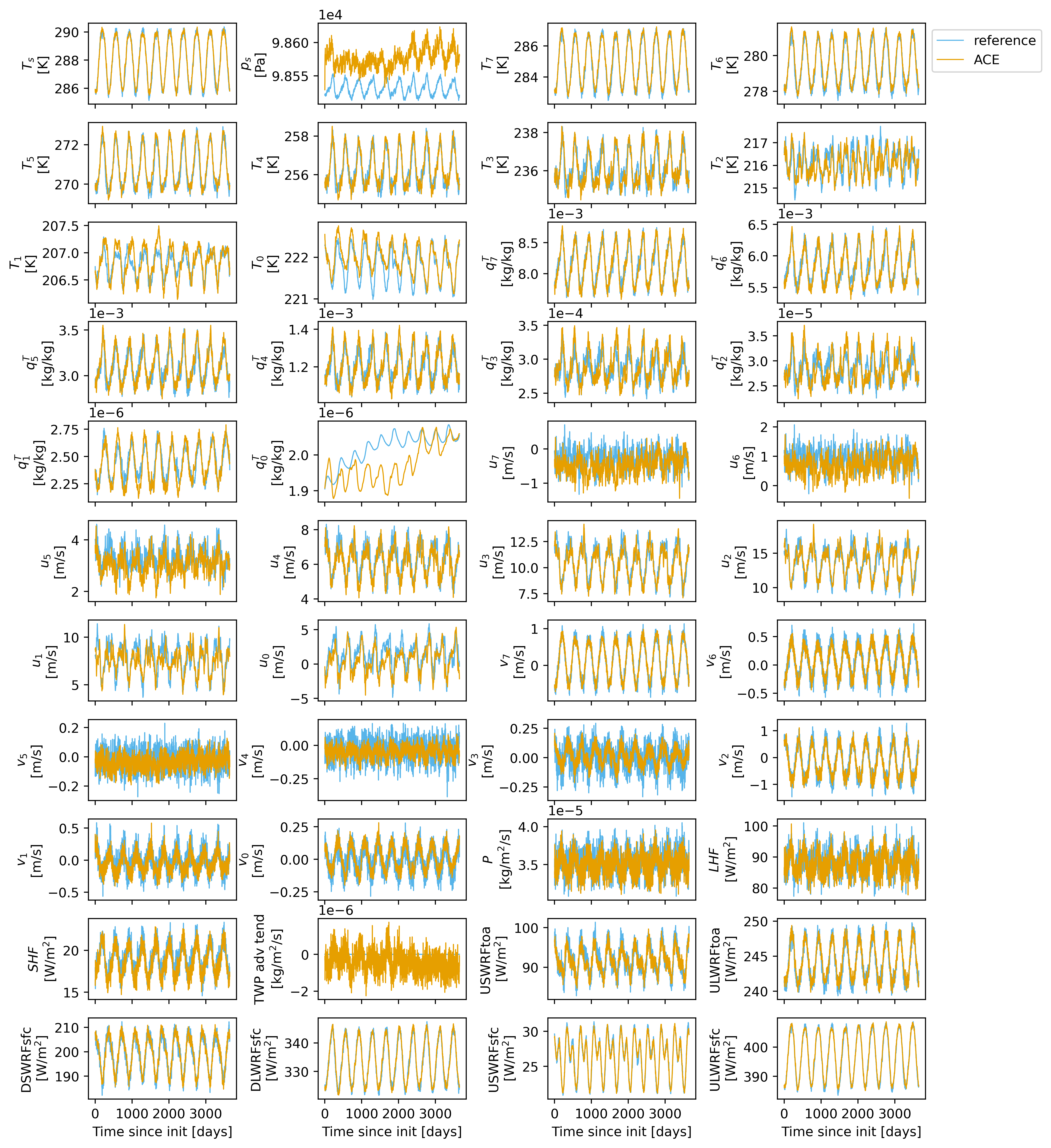}
  \caption{The daily averaged global mean (Equation~\ref{eq:globalmean}) of all output variables for ACE and the reference simulation.}
  \label{fig:globaltimeseriesall}
\end{figure}

Figure~\ref{fig:globaltimeseriesall} shows the timeseries of the global mean for all output variables. Most variables track the global-mean season cycle well and none have large drifts that persist throughout the 10-year simulation. Nevertheless, there are deviations from the expected global mean that are larger than interannual variability (specifically in $p_s$, $q_0^T$ and a number of the northward wind $v_k$ variables).

\section{100-year forecasts}
\begin{figure}
  \centering
  \includegraphics[width=\textwidth]{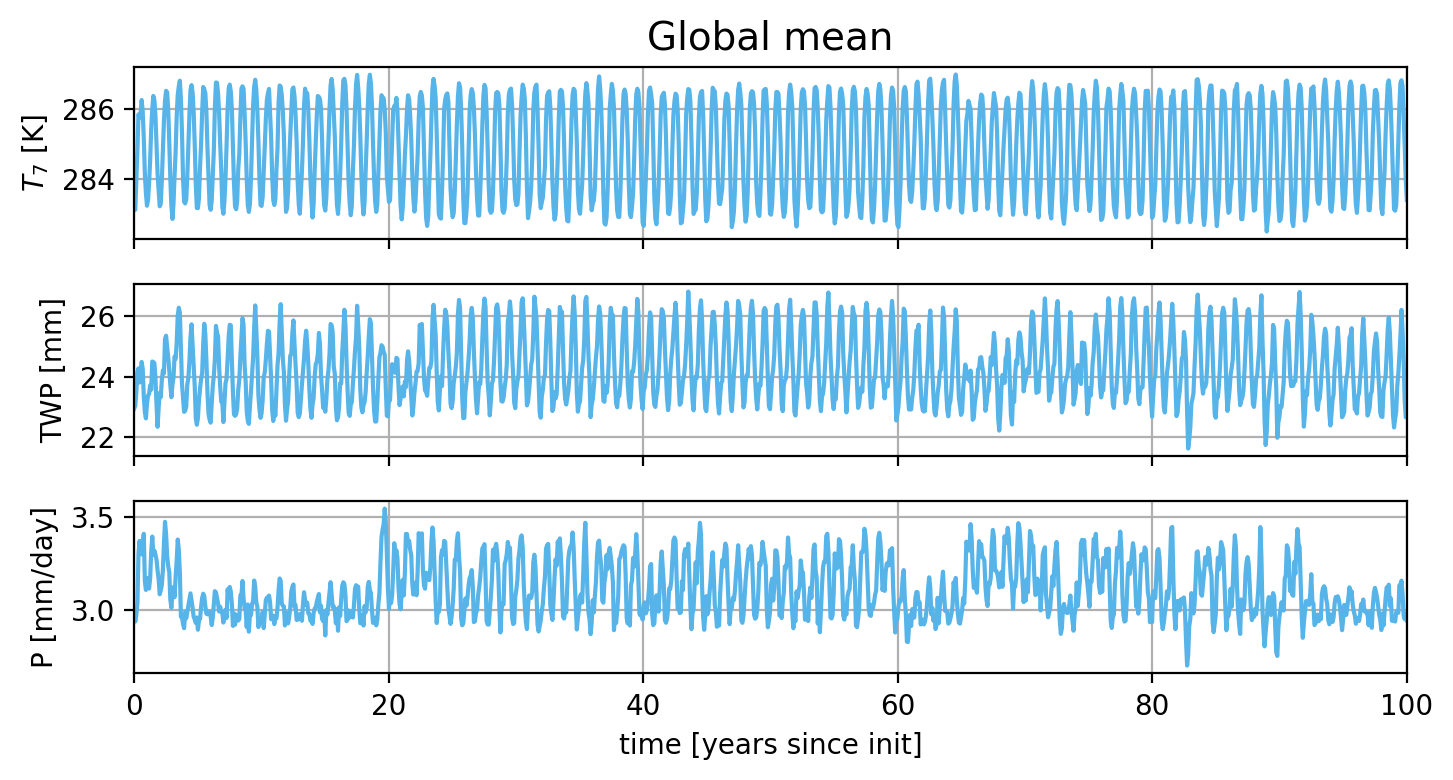}
  \caption{Timeseries of monthly-mean and global-mean (top) $T_7$, (middle) total water path and (bottom) surface precipitation rate $P$ from 100-year forecast.}
  \label{fig:100yearrun}
\end{figure}

Since our forcing dataset is annually repeating, it is straightforward to repeat it in order to allow arbitrarily long forecasts. We used this feature to do a 100-year long forecast. The forecast was stable and no variables showed long-term drifts. However, there are some unrealistic fluctuations on annual timescales, for example for global mean surface precipitation rate (bottom row of Figure~\ref{fig:100yearrun}).

\section{Global mean conservation}
\label{appendix:globalbudgets}
\begin{figure}
  \centering
  \includegraphics[width=\textwidth]{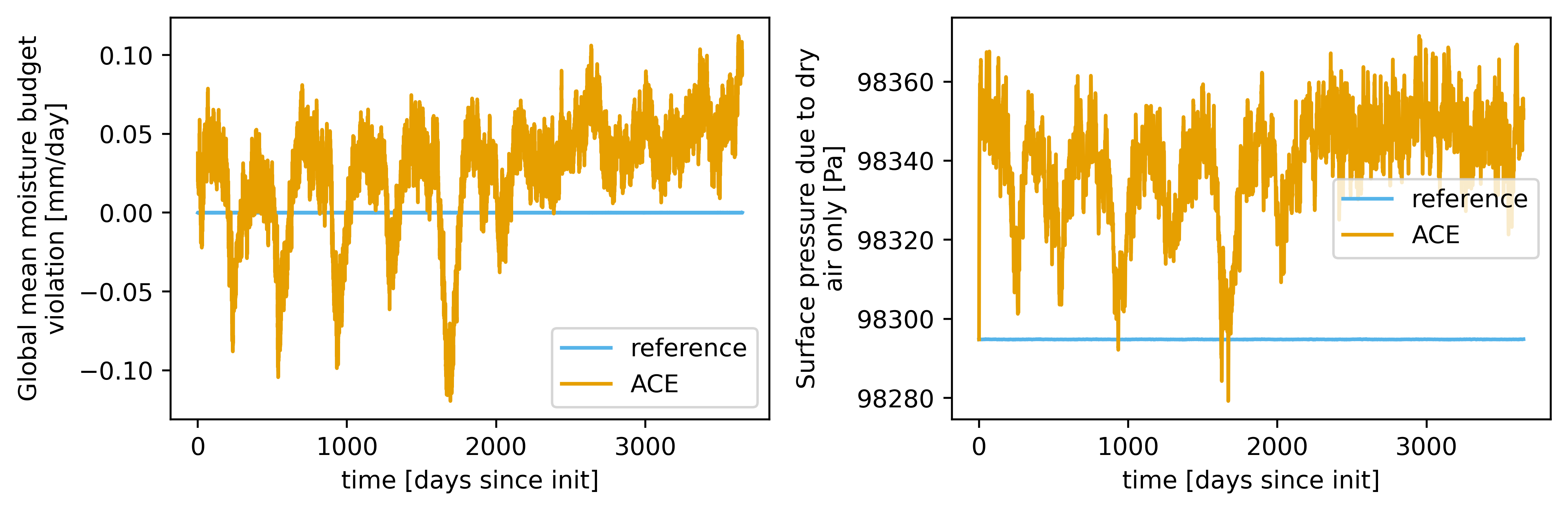}
  \caption{Left: timeseries of the global mean moisture budget violation (LHS minus RHS of Equation~\ref{eq:twp}). Right: timeseries of global mean surface pressure due to dry air only (Equation~\ref{eq:dryair}).}
  \label{fig:globalbudget}
\end{figure}

The last paragraph of Section~\ref{sec:results} discussed the column budget of moisture and showed that ACE nearly conserves column moisture for individual timesteps. Here we compute the area-weighted global mean of Equation~\ref{eq:twp} and show the global budget violation (left-hand side of Equation~\ref{eq:twp} minus right-hand side of Equation~\ref{eq:twp}) in the left panel of Figure~\ref{fig:globalbudget}. Although for most of the simulation the violation is not overly large (magnitude less than 0.1 mm/day; compared to typical values of global mean precipitation or evaporation of 3 mm/day) at certain times in the simulation the global budget violation is up to magnitude 0.4 mm/day.

We also consider the conservation of global dry air mass. Surface pressure due to dry air only is
\begin{equation}
\label{eq:dryair}
  p_s^{dry} = p_s - g \cdot TWP.
\end{equation}
On the timescales we are considering, there is no flux of dry air mass through the top of atmosphere or through the Earth's surface. Therefore, the global mean surface pressure due to dry air should remain exactly constant. Indeed this is the case for the reference model. The right panel of Figure \ref{fig:globalbudget} shows slight deviations for ACE of magnitude up to 150 Pa, but no systematic multi-year drift.

It is possible that adding terms to the loss function to encourage global mean conservation of moisture and dry air mass \cite{Raissi2019} would improve these characteristics for long simulations.

\section{Normalization Strategy}
\label{appendix:normalization}

This section describes the ``residual'' scaling strategy and compares it to the ``full-field'' normalization done in \cite{FourCastNet}.
First the standard scaling mean and standard deviations are computed over time, latitude and longitude, without area weighting:
\begin{align}
  \mu(a) &= \underset{t, \phi, \lambda}{\mathrm{mean}} \, a(t, \phi, \lambda) \\
  \sigma(a) &= \underset{t, \phi, \lambda}{\mathrm{std}} a(t, \phi, \lambda).
\end{align}
The ``full-field'' normalization (as in \cite{FourCastNet}) is defined as $a_{ff}=(a-\mu(a))/\sigma(a)$. Some prognostic variables, for example surface pressure $p_s$, vary more strongly over space than over time, so their normalized timestep-to-timestep differences $a_{ff}' = a_{ff}(t+1) - a_{ff}(t)$ (which is what we'd really like to get right for each prognostic variable) vary strongly between variables. It is cumbersome to change the loss function in FourCastNet to reflect this goal. As a pragmatic substitute, we modify the normalization strategy to give equal contributions from tendency errors in each variable to the loss function
(see also Section 3.3.3 of \cite{Keisler2022} which normalized the loss with a similar motivation). 
First we compute the forward increment of the normalized variables as:
\begin{equation}
\label{eq:affprime}
  a_{ff}'(t) = a_{ff}(t+1) - a_{ff}(t).
\end{equation}
Define $\sigma(a_{ff}')$ is the standard deviation of $a_{ff}'$.  Neglecting any temporal trend in $a_{ff}$ across the training period, $\mu(a_{ff}')=0$, so
\begin{equation}\label{eq:residualvariance}  \sigma^2(a_{ff}') = \frac{1}{TXY}{\sum_{t,x,y}(a_{ff}')^2} 
\end{equation}
We rescale the full-field-normalized variables as follows:  
\begin{eqnarray}
  a_{res} &=& a_{ff} / \sigma_{res}(a) \\
 \sigma_{res}(a) &=&  {\sigma(a_{ff}')}/{\overline{\sigma(a_{ff}')}^{g}}
\end{eqnarray}
This ``residual'' rescaling includes a factor  $\overline{\sigma(a_{ff}')}^{g}$, the geometric mean of $\sigma(a_{ff}')$ across all variables, so that the standard deviations of the rescaled variables stay scattered about 1. 

\begin{figure}
  \centering
  \includegraphics[width=0.7\textwidth]{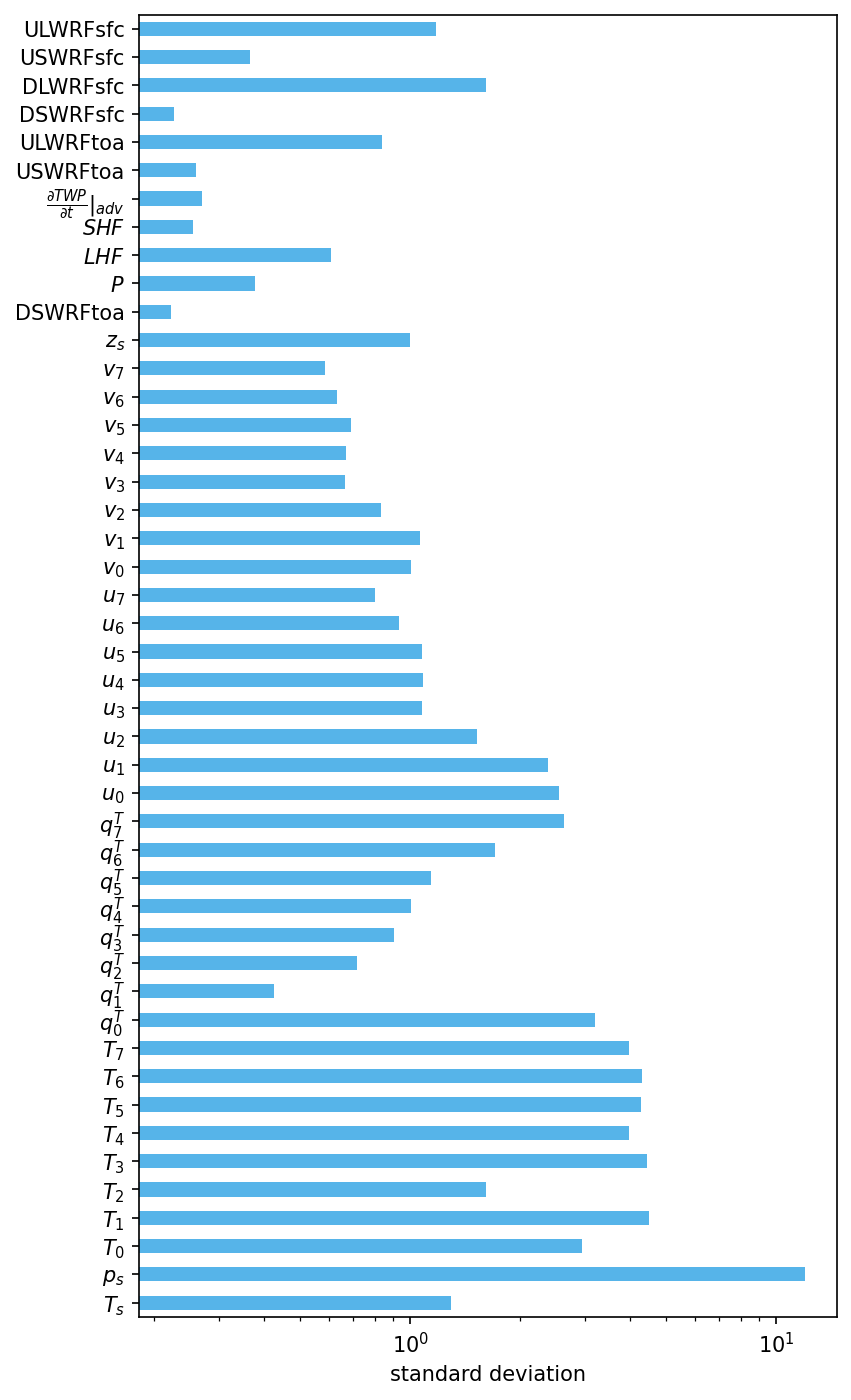}
  \caption{The standard deviation of normalized variables using the ``residual'' scaling strategy. Table~\ref{table:variables} gives meaning of each variable.}
  \label{fig:normstddev}
\end{figure}

Figure~\ref{fig:normstddev} shows the ``residual'' rescaling factors $\sigma_{res}(a)$ of all variables $a$. The range is about a factor of 30, with the surface pressure being up-weighted the most relative to the ``full field'' normalization. 
Residual normalization is more justifiable for prognostic variables (which are both inputs and outputs) than for diagnostic variables. Hence, we attempted to apply residual scaling only to prognostic variables, but this leads to slightly larger time-mean biases than also rescaling diagnostic variables.

\section{Ablations}
\subsection{Normalization}
\label{appendix:normalizationablation}
\begin{figure}
  \centering
  \includegraphics[width=\textwidth]{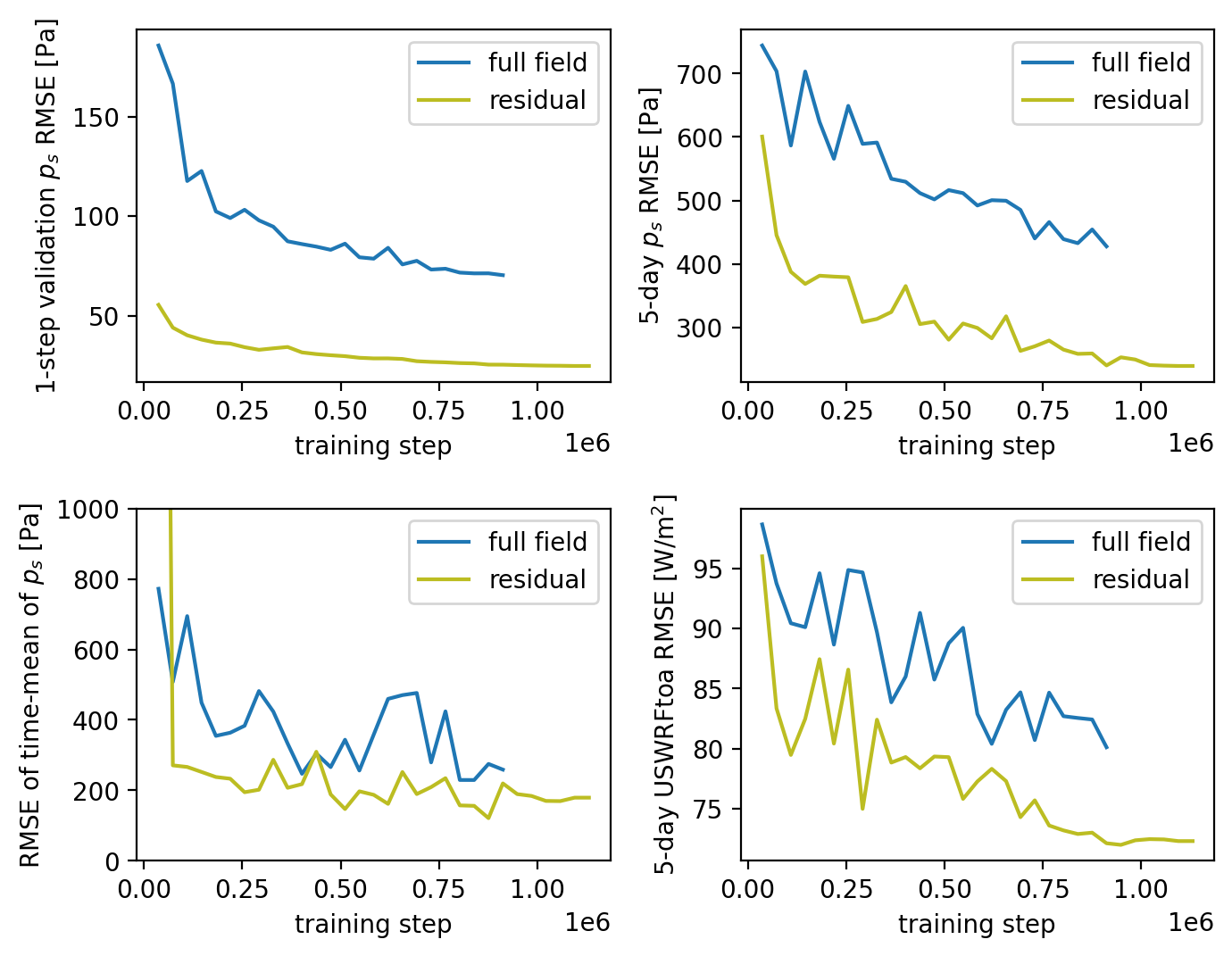}
  \caption{Four metrics evaluated on models which use (blue) full field normalization and (olive) residual normalization, shown as a function of training step and reported once per epoch. The metrics are (top-left) RMSE of 1-step $p_s$ prediction on validation data, (top-right) RMSE of $p_s$ after 5 days of autoregressive inference, (bottom-left) RMSE of time-mean $p_s$ over 1-year simulation and (bottom-right) RMSE of USWRFtoa after 5 days of autoregressive inference. }
  \label{fig:normablate}
\end{figure}

Note that neither of the models used in this ablation (Figure~\ref{fig:normablate}) have the same configuration as the for the results shown in the main body of the manuscript. In particular, neither of them use the surface type fractions as inputs.

To show the impact of the ``residual scaling'' normalization choice described in Appendix~\ref{appendix:normalization}, Figure~\ref{fig:normablate} shows the 1-step validation RMSE, 5-day RMSE and time-mean RMSE ($RMSE\textnormal{-}TM_{p_s}$) of surface pressure and 5-day RMSE of USWRFtoa as a function of training steps. Not surprisingly since we are weighting it more strongly, a large (about 50\%) decrease in 1-step RMSE is seen for $p_s$. We also see a large improvement for 5-day RMSE, and smaller improvement for time-mean RMSE of $p_s$. Notably, even though the USWRFtoa variable is weighted less heavily in the loss function (see Figure~\ref{fig:normstddev}) its 5-day RMSE is substantially improved with residual scaling. We believe this is because the improved autoregressive performance of $p_s$ leads to more accurate predictions for other variables also.

\subsection{Dataset size}
To show importance of dataset, here we compare training curves for two experiments which are similarly configured. The only difference is that the first uses a dataset with 10-years of training data (14607 samples; a single member of the 10-member ensemble of reference simulations) and trains for 75 epochs while the second uses a 100-year dataset (146070 samples) and trains for 15 epochs. Note the latter training is not the same as the one described in the main text above. It does not use surface height or surface type fractions as inputs and it uses a 5x larger initial learning rate.

\begin{figure}
  \centering
  \includegraphics[width=0.7\textwidth]{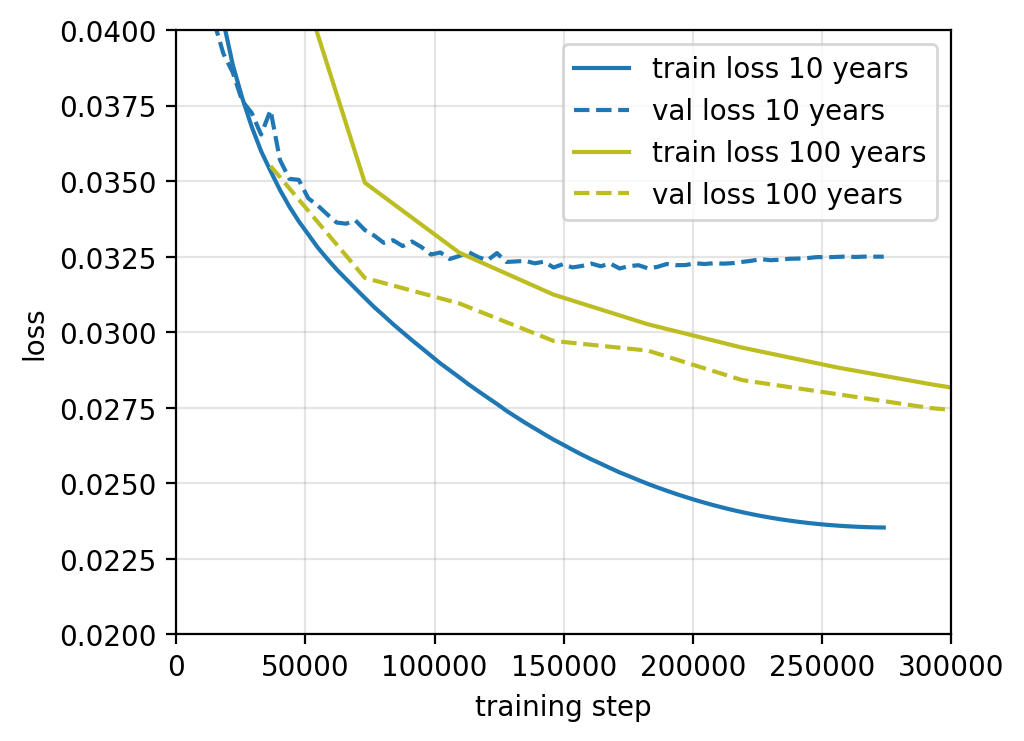}
  \caption{The (solid) train loss and (dashed) validation loss for training runs on (blue) a 10-year dataset and (olive) a 100-year dataset.}
  \label{fig:loss}
\end{figure}

Figure~\ref{fig:loss} shows the training and validation loss for the models trained on the two datasets. The validation loss here is computed on 80 samples from the independent validation dataset. The model trained on only 10 years of data shows clear signs of overfitting that does not occur for the 100-year dataset.

\begin{figure}
  \centering
  \includegraphics[width=\textwidth]{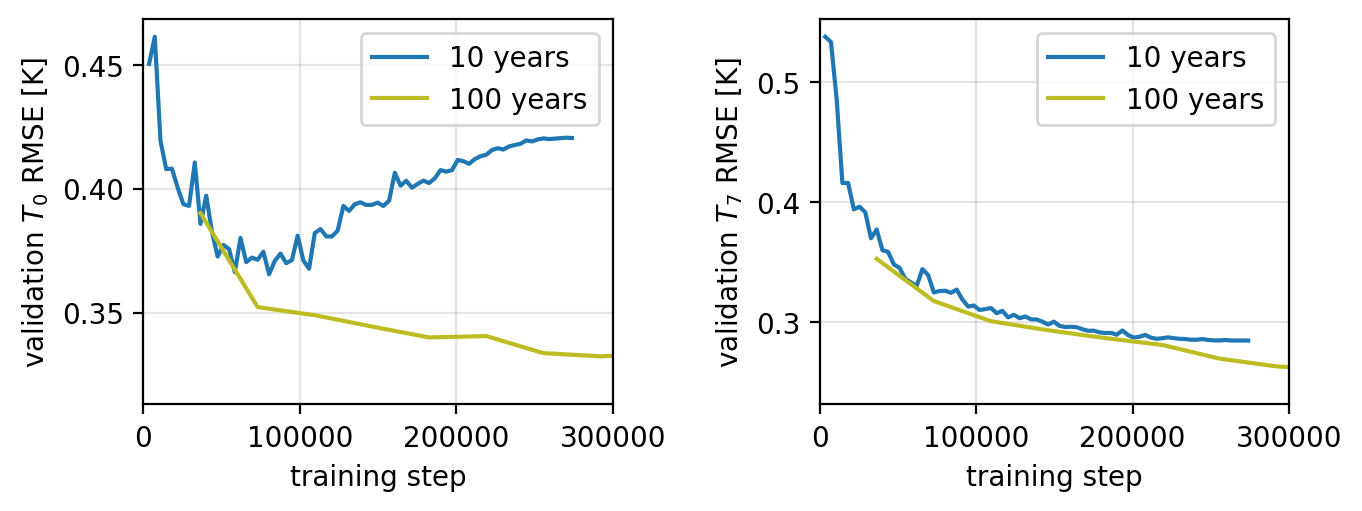}
  \caption{The validation RMSE of (left) $T_0$ and (right) $T_7$ for models trained on (blue) a 10-year dataset and (olive) a 100-year dataset.}
  \label{fig:valrmse}
\end{figure}

When examining the variable-by-variable validation RMSE, we find that only stratospheric variables overfit for the 10-year dataset. As an example, Figure~\ref{fig:valrmse} compares the validation RMSE for the upper stratospheric $T_0$ and the near-surface $T_7$. This makes it apparent that the overfitting is happening for temperature in the stratosphere. This is also the case for other stratospheric variables $T_1$, $v_0$, $u_0$, $q^T_0$ and $q^T_1$ (not shown). Given that stratospheric dynamics are typically slower than variability in the troposphere, it is not a surprise that overfitting is a particular concern in the uppermost model layers. Interestingly, ML-based weather prediction systems have reported relatively poor performance in the stratosphere (e.g.\ Fig.\ 2 of \cite{GraphCast}).

\end{document}